\documentstyle[11pt]{article}

\newcommand{\hf}{\frac{1}{2}}

\newcommand{\xn}{x_{n}}

\newcommand{\e}{e^{i kX(z)}}

\newcommand{\kim}{ k_{1}^{\mu}}                                      
\newcommand{\kom}{ k_{0}^{\mu}}                                      
\newcommand{\ki}{ k_{1}}
\newcommand{\yn}{ Y_{n}}

\newcommand{\ko}{ k_{0}}

\newcommand{\kin}{ k_{1}^{\nu}}                                      
\newcommand{\kon}{ k_{0}^{\nu}}

\newcommand{\gvks}{ e^{i\sum _{n\ge 0 }\int dt ~k_{n}(t )Y_{n}(t )}} 
\newcommand{\GVKS}{e^{i\sum _{n \ge 0 }\int dt' ~k_{n}(t ')\tY _{n}
(t ')}}

\newcommand{\dsi}{\frac{\partial}{\partial x_{1}}}

\newcommand{\dsn}{\frac{\partial }{\partial x_{n}}}  
\newcommand{\dsm}{\frac{\partial }{\partial x_{m}}}  
         
\newcommand{\dsq}{\frac{\partial }{\partial x_{n+m}}}

\newcommand{\dsnm}{\frac{\partial ^{2}}{\partial x_{n}\partial x_{m}}}

\newcommand{\p}{\partial}                                           
                                           
\newcommand{\pp}{\partial ^{2}}

\newcommand{\al}{\alpha }                                             
\newcommand{\aln}{\alpha _{n}} 
\newcommand{\tY}{\tilde Y}

\newcommand{\la}{\mbox{$ \lambda $}} 
\newcommand{\be}{\begin{equation}}
\newcommand{\br}{\begin{eqnarray}}
\newcommand{\ee}{\end{equation}} 
\newcommand{\er}{\end{eqnarray}}

\newcommand{\ppp}{\mbox {$ \partial ^{3}$}}

\begin{document}
\renewcommand{\theequation}{\thesubsection.\arabic{equation}}

\title{
\hfill\parbox{4cm}{\normalsize IMSC/2007/05/06\\
                               hep-th/yymmnnn}\\        
\vspace{2cm}
Gauge Invariant Exact Renormalization Group and Perfect Actions in the Open Bosonic
String Theory.
\author{B. Sathiapalan\\ {\em Institute of Mathematical Sciences}\\
{\em Taramani}\\{\em Chennai, India 600113}\\ bala@imsc.res.in}}           
\maketitle       

\begin{abstract} 
 The exact renormalization group is applied to the world
sheet theory describing bosonic open string backgrounds to obtain the equations of motion
for the fields of the open string. Using loop variable techniques the equations can be
constructed to be gauge invariant. Furthermore they are valid off the (free) mass shell.
This requires keeping a finite cutoff. Thus we have the interesting situation of a scale
invariant world sheet theory with a finite world sheet cutoff. This is possible because
there are an infinite number of operators whose coefficients can be tuned.  This is in the same sense that "perfect actions" or "improved actions" have been proposed in lattice gauge theory to reproduce the continuum results even while keeping a finite lattice spacing.  

\end{abstract}

\newpage

\section{Introduction}
The renormalization group has been applied to the world sheet action for a string propagating in non-trivial backgrounds to obtain equations of motion. [\cite{L}-\cite{T}]. As a  generalization of this technique, 
Loop Variable techniques have been used to write down gauge invariant equations
of motion for both open and closed strings \cite{BSLV,BSREV,BSCS}.  These are essentially equations that set to zero the change in coupling constants of the two dimensional world sheet field under scale  transformations, i.e these are
conditions for a fixed point under a renormalization group (RG) transformation. 
There are a couple of noteworthy features :
 One is that gauge invariance (in space time)  necessitates including all the modes of the string.
 Another is that in order to deal with non-marginal vertex operators, i.e.
 for  space time fields that do not obey the mass shell constraint, it is necessary to keep the world sheet cutoff
finite, at least in the intermediate stages of the calculation. Some aspects of the finite
cutoff theory has been discussed  in \cite{BSPT, BSFC,BSLC} where it was shown that
if one keeps a finite cutoff, the proper time equation for the tachyon (which in this situation is related to the RG equation),  become quadratic. This is as expected both from
string field theory and also from the exact renormalization group [\cite{WK}-\cite{P}]. In \cite{BSLC} it was also shown that one can make precise contact with light cone string field theory by keeping a finite cutoff \footnote{In light cone
string field theory there is also a quartic term, which is a subtlety that is not addressed here.}  In \cite{BSFC} 
it was also shown that if one wants to maintain gauge invariance
while maintaining a finite cutoff one needs to include all the massive modes in the proper time equation.
In this sense string field theory \cite{WS,SZ,Wi2,BZ}can be thought of as a way
of keeping a finite cutoff while maintaining gauge invariance. 
 
Another approach to off shell string theory
is the background independent approach pioneered in \cite{Wi} and further developed in \cite{LW,Sh,KMM}.
The connection with the RG approach is discussed in \cite{Sh,KMM}.

Implicit in the RG approach is the interesting fact that one is maintaining a finite cutoff while discussing a scale
invariant theory. This is possible because one has an action with all possible operators and thus it
is conceivable that with an infinite number of fine tunings one can satisfy the fixed point conditions 
and attain scale invariance even when the cutoff is non zero.

The main aim of this work is to write the gauge invariant loop variable equations in the form of an exact
RG  (ERG) equation (with finite cutoff). (See \cite{P,BB1,BB2,S1,IIS} for a discussion of many conceptual issues encountered in the ERG.)
The gauge fixed version is quadratic in fields  and therefore is similar to string field theory. 
 But in the absence
of a world sheet symmetry principle there could be  a lot of arbitrariness in the scheme and one is not sure
if the result is equivalent to string theory, in particular whether it can be made gauge invariant. In the loop variable approach, space-time gauge invariance is built in. This ensures that in the critical dimension
negative norm states decouple.
Having space time gauge invariance  built into it at the outset is thus reassuring.  The fact that we do not rely on world sheet reparametrization (or BRS) for space time gauge invariance
is an advantage in that  there is no clash at any time between world sheet regularization and  
spacetime gauge invariance.  This gives a lot of freedom in choice of regularization. This was
exploited recently to construct a (free) higher spin action in AdS space-time \cite{BSA}.  This is not so straightforward 
in string field theory.  However unlike string field theory the equations
of motion  are  not quadratic - they involve higher order terms.

The phenomenon of scale invariance at finite cutoff is interesting in its own right. This is related to the
idea of "improved actions" \cite{KS} or "perfect actions" \cite{PH} introduced in the context of lattice
guage theory. (See also \cite{BB1,BB2}.) The basic idea is that if one is exactly on the RG trajectory connecting the UV and IR fixed
points then one is infinitely far from the continuum (i.e. one has a finite lattice spacing) and yet it
is physically equivalent to it - because it is on the same RG trajectory.  One can take this one step
further and say that if one starts at the fixed point itself with a finite cutoff then after an infinite
number of steps, when one has reached  the continuum, we still have the same action! Thus the fixed point action
is scale invariant with either finite or zero cutoff.  

In string theory this perfect action also describes
the precise values of all the infinite number of massive modes in a background that is a solution to the classical equations of motion. The extension to the quantum theory (i.e loop corrections) is an open question.

This paper is organized as follows: In Section 2 we derive the ERG in position space. In Section 3 we apply
it to gauge fixed backgrounds - as one would in the "old covariant formulation" of string theory. In section 4
we derive the gauge invariant version using loop variables. Section 5 contains some conclusions and 
speculations.

\section{RG in Position Space}

In this section we derive the exact RG in position space. This is essentially a repetition of Wilson's
original derivation \cite{WK}.  We include it here only because usual discussions use momentum space rather than position space.  We start with  point particle quantum mechanics:

\subsection{Quantum Mechanics}

We start with the Schrodinger equation
\be 
i\frac{\partial \psi}{\p t} =-\frac{\pp \psi }{\p y^2}
\ee  
for which the Green's function is $\frac{1}{\sqrt{2\pi (t_2-t_1)}}e^{i\frac{(y_2-y_1)^2}{2(t_2-t_1)}}$,
and change variables :$y = xe^\tau , it = e^{2\tau}$ and $ \psi ' = e^\tau \psi$ to get the
differential equation
\be
\frac{\p \psi '}{\p \tau} = \frac{\p}{\p x }(\frac{\p}{\p x} + x ) \psi ' 
\ee
with Green's function
\be
G(x_2, \tau _2; x_1 , 0) = \frac{1}{\sqrt{2\pi (1- e^ {-2\tau _2})}}
e ^{- \frac{(x_2 - x_1 e^{-\tau _2)^2}} {2(1-e^{-2\tau _2 )}}}
\ee

Thus as $\tau _2 \rightarrow \infty $ it goes over to $\frac{1}{\sqrt{2\pi}}e^{-\hf x_2^2}$. As $\tau _2 \rightarrow 0$ it goes to $\delta (x_1-x_2 )$. 
\[ \psi (x_2, \tau _2 ) = \int dx_1 G (x_2, \tau _2; x_1,0 )\psi (x_1, 0)
\]

So $\psi (x_2, \tau _2 )$ goes from being unintegrated
$\psi (x_1)$ to  completely integrated $\frac{1}{\sqrt{2\pi}}e^{-\hf x_2^2} \int dx_1 \psi (x_1)$. 
Thus consider
\be
 \frac{\p}{\p\tau } \psi (x_2, \tau) =  \frac{\p}{\p x_2}(\frac{\p}{\p x_2} + x_2) \psi (x_2, \tau )
\ee
with initial condition $\psi (x, 0) $
Thus if we define $Z(\tau ) = \int d x_2 \psi (x_2, \tau )$, where $\psi $ obeys the above equation, we see that $\frac{d}{d\tau} Z =0$. Also for $\tau =0$ $\psi$ is the unintegrated $\psi (x,0)$. At $\tau = \infty$ it
is proportional to the integrated object $\int dx \psi (x,0)$. $Z(\tau )$ has the same value. Thus as
$\tau$ increases the integrand in $Z$ is more completely integrated.

We need to repeat this for the case where the initial wave function is replaced by $e^{\frac{i}{\hbar}S[x ]}$ where
$x$ denotes the space-time coordinates. Then for $\tau = \infty$ $\psi  \approx \int {\cal D }x e^{iS[x]}$ the integrated partition function. At $\tau =0$ it is the unintegrated $e^{iS[x]}$. $Z (\tau )$ is the fully integrated
partition function for all $\tau$. We shall also split the action into a kinetic term and interaction term as in \cite{P}.
Thus in the quantum mechanical case discussed above we write $\psi = e ^{-\hf x^2  f(\tau ) +L(x)}$

By choosing $a,b,B$ suitably  ( $b=2af , B = \frac{\dot f}{bf} $) in 
\[ \frac{\p \psi}{\p \tau} = B \frac{\p}{\p x} ( a \frac{\p}{\p x} + bx) \psi (x,\tau )
\] we get 
\be
\frac{\p L}{\p \tau} = \frac{\dot f}{2 f^2}[\frac{\pp L}{\p x^2} + (\frac{\p L}{\p x})^2 ]
\ee
Note that if $f = G^{-1}$ ($G$ is like the propagator) then $\frac{\dot f}{f^2} = -\dot G$

\subsection{Field Theory}
We now apply this to a Euclidean field theory.

\be
\psi = e^{-\hf \int dz \int dz' X(z) G^{-1} (z,z') X(z') + \int dz L[X(z)]}
\ee

We apply the operator
\be   \label{27}
\int dz \int dz' B(z,z') \frac{\delta}{\delta X(z')} [\frac{\delta}{\delta X(z)} + \int b(z,z'')X(z'') ] 
\ee
to $\psi$ and require, as before, that this should be equal to $\frac{\p \psi}{\p \tau}$. 

We get the following five terms (all multiplied by $B$:
\[ (b-G^{-1})(z,z')
\]
\[  + [\frac{\pp L}{\p X(z) \p X(z')} \delta (z-z') + \frac{\p L}{\p X(z)} \frac{\p L}{\p X(z')}]
\]
\[   +\frac{\p L}{\p X(z)} (- \int G^{-1} (z',z'') X(z'') dz'')
\]
\[  +\frac{\p L}{\p X(z')} ( \int (b-G^{-1}) (z,z'') X(z'') dz'')
\]
\be - [(b- G^{-1}) X](z) [G^{-1} X](z') 
\ee
The first term is independent of $X$ and is therefore an unimportant overall constant. If we choose
$b = 2G^{-1}$, the third and fourth terms add up to zero. 

Thus the second  term becomes
\be 
\int dz \int dz' B(z,z')  [\frac{\pp L}{\p X(z) \p X(z')} \delta (z-z') + \frac{\p L}{\p X(z)} \frac{\p L}{\p X(z')}]
\ee

and the last term becomes:
\be
-\int dz \int dz' B(z,z') dz'' dz''' G^{-1} (z,z'') X(z'') G^{-1} (z', z''') X(z''') \psi
\ee

We can set (\ref{27})  equal to \[\frac{\p \psi}{\p \tau} = -\hf \int dz \int dz' X(z) \frac{\p G^{-1}}{\p \tau} (z,z') X(z') \psi
+ \int dz \frac{\p L}{\p \tau} \psi .\]
This ensures that $Z= \int {\cal D} X \psi $ satisfies $\frac{\p Z}{\p \tau} =0$.
If we now set $B = -\hf \dot G^{-1} (z,z')$ the equation for $\psi$ reduces to:
\be	\label{ERG}
\int dz \frac{\p L}{\p \tau} = -\int dz \int dz' \hf \dot G (z,z') [\frac{\pp L}{\p X(z) \p X(z')} \delta (z-z') + \frac{\p L}{\p X(z)} \frac{\p L}{\p X(z')}]
\ee
If we now interpret $\tau$ as $ln~a$ this becomes easy to interpret as an RG equation diagrammatically
as done in \cite{P}:  the first term in the RHS represents
contractions of fields at the same point - self contractions within an operator, and the second one represents contractions between fields at two different points - between two different operators. 

\section{ERG in the Old Covariant Formalism}

We can assume that there is an infrared cutoff in all the integrals i.e. $\int _{-R}^R dz$ - otherwise in a
conformal field theory there could be infrared divergences in the integrals. When we integrate modes above 
a value $\Lambda = {1\over a}$ analyticity would demand that we also partially integrate some of the 
low energy modes. This is a potential source of IR divergences and could bring in dependences on the parameter  $R\over a$. However if the cutoff is sharp enough (consistent with analyticity) one can safely take the
limit $R \rightarrow \infty$. It is also possible to have a  cutoff so sharp that even for
finite $R$ the ERG equations have no dependence on $R\over a$. However such a  cutoff would
not be consistent with analyticity. Analyticity is important in the present case because we will
be making essential use of the OPE to reexpress non-local products of operators as higher dimensional
local terms in the action.  Since our starting point is an action that contains all the open
string modes as backgrounds, this is a perfectly reasonable thing to do.    Thus we can assume that the action is
\[
S = \int _0 ^R dz L[X(z)] = \int _0^Rdz \int dk [{\phi (k) \over a}e^{ikX(z)} + A_\mu (k) \p _z X^\mu e^{ikX(z)}
\]
\be \label{S}
+
\hf a S_2(k)^\mu \pp _z X^\mu e^{ikX(z)} + a S^{\mu \nu}(k)\p X^\mu \p X^\nu e^{ikX(z)}+...]
\ee

Before we implement the ERG we need a specific form for $\dot G(z,z')=\dot G(z-z')$.
As mentioned above we need $\dot G (u)$ to be short ranged, otherwise the dimensionless ratio $R\over a$ is bound to enter
in the equations. 

One can make use of functions of the form $e^{-{1\over x^2}}\theta (x)$ that vanishes at $x\leq 0$ along with all
derivatives and yet is continuous at $x=0$ along with all derivatives.  The precise
form is not very important - although it will fix the various numerical constants in  the RHS of the ERG. The main property is that it should vanish for $|u| > a$. Thus it could be 
$e^{-\frac{1}{(u-a)^2}}e^{-\frac{1}{(u+a)^2}}e^{\frac{2}{a^2}}$ for $|u| \leq a$.  And   it has the usual form - $G(u) = ln~u$ for $|u| >a$ so that $\dot G(u) =0$. With this function it is easy to see that even
for finite $R$ the equations do not depend on $R\over a$. However being non-analytic one cannot perform
an OPE - at least not in the usual way that involves Taylor expansions. 

We will use a different cutoff Green's function:
\be G(u) = \int {d^2k\over (2\pi )^2} \frac{e^{iku}e^{ - a^2 k^2} }{k^2}
\ee
This has a cutoff at short distances of $O(a)$ and at long distances reduces to the usual propagator.
We now apply the ERG (\ref{ERG}) to the action S (\ref{S}).

The LHS gives
\be  \label{LHS}
\int dz \int dk ~[ {\beta _{\phi (k)} - \phi (k) \over a} \e + \beta _{A^\mu(k)} \p _z X^\mu(z)\e +...]  
\ee
where $\beta _g \equiv \dot g$.
The first term of the RHS gives
\be   \label{RHSI}
\int dz \int dk \hf (-k^2) {\e \over a} \phi (k)
\ee

The second term gives
\be   \label{RHSII}
\int dk_1 \int dk_2 {\phi (k_1) \phi (k_2) \over a^2} ({-k_1.k_2 \over 2}) \int _{-R}^{+R} du \dot G(u) e^{ik_1.X(z)}e^{ik_2.X(z+u)}
\ee
One can do an OPE for the product of exponentials to get
\[
e^{i(k_1+k_2)X(z) + ik_1 [u\p _z X + {u^2\over 2} \pp X + ...]}
\]

This gives 
\[
e^{i(k_1+k_2)X(z)} \int _{-R}^{+R} du ~\dot G(u) + ik_1 \p X e^{i(k_1+k_2)X(z)} \int _{-R}^{+R} du~ u \dot G(u)
\]
\[+ i k_1 {\pp X \over 2}  e^{i(k_1+k_2)X(z)} \int _{-R}^{+R} du ~ u^2 \dot G(u) +
{i k _\mu ik_ \nu \over 2} \p X^\mu \p X^\nu e^{i(k_1+k_2)X(z)} \int _{-R}^{+R} du~ u^2 \dot G(u) 
\]
It is easy to see that the first term of the OPE contributes to the tachyon equation:
\be
\beta _{\phi (k)} -\phi (k) = \phi (k) ({-k^2 \over 2}) - 
\hf \int dk_1 \phi (k_1) \phi (k-k_1) {k_1.(k-k_1)\over 2a}\int _{-R}^{+R} du~\dot G(u) 
\ee

Similarly the second term of the OPE contributes to the photon equation:
\be
\beta _{A^\mu (k)} =\int dk_1 {\phi (k_1) \phi (k-k_1) \over a^2} ({-k_1(.k-k_1) \over 2}) ik_1 ^\mu \int _{-R}^{+R} du~ u \dot G(u)
\ee

We have thus obtained the contribution of the tachyon field to the beta functions of the tachyon and photon.
Similarly there ae contributions to the beta functions of the higher spin massive fields $S_\mu, S_{\mu \nu}$ etc.
 Note that the dimensionless number $R/a$ does appear as expected because of the analytic nature of the cutoff. 
However since there are no infrared divergences one can take the limit $R\rightarrow 0$ without any
problem. $\dot G(u) = \frac{1}{\pi}e^{-\frac{u^2}{4a^2}}$.  Thus integrals $\int _{-R}^Rdu \dot G(u) u^n$
all have  $R/a$-dependent pieces that contain the factor $e^{-\frac{R^2}{a^2}}$. So in the $R\rightarrow \infty$
limit, all $R/a$ dependence disappears, and the equations become completely independent of $a$. \footnote{This can be understood as an example of finite size scaling, which has been much studied \cite{JC}.}
Thus
the conditions for the fixed point do not depend on $a$, i.e. as mentioned in the introducton, there is  scale invariance even though the lattice spacing $a$ is non zero. This is the kind of situation envisaged
in \cite{KS, PH} where the coupling constants of the irrelevant operators are all tuned so that
the physical quantities calculated with this action do not depend on the cutoff $a$, and thus they have
the same values as in the continuum.  These are the "improved" actions \cite{KS} or "perfect" actions \cite{PH}.
If on top of that, the background fields are tuned to satisfy the fixed point condition,
 then we have a scale invariant theory, even while the lattice spacing is non-zero.

One can also include the contribution due to the photon field in the RHS as shown below:
\[
\frac{\delta}{\delta X^\mu (z)} \int dz''~\int dk~ A_\nu (k)\p _{z''}X^\nu (z'') e^{ikX(z'')}
\]
\[
= \int dz'' \int dk A_\nu (k) [ \delta ^{\mu \nu} \p _{z''} \delta (z-z'') e^{ikX(z'')} + \p _{z''} X^\nu (z'')ik^\mu \delta (z-z'') \e ]
\]
\[ =\int dz'' \int dk A_\nu (k) [ -\delta ^{\mu \nu}  \delta (z-z'')ik^\rho \p _{z''}X^\rho e^{ikX(z'')} + \p _{z''} X^\nu (z'')ik^\mu \delta (z-z'') \e
\]
\[
=\int dz'' \int dk [-A^\mu (k)  ik^\nu  + ik^\mu A^\nu ] \p _{z''} X^\nu (z'')\delta (z-z'') \e
\]
\[
\frac{\delta ^2}{\delta X^\mu (z')X^\mu (z)}\int dz~L = \int dz'' \int dk \delta (z-z'') [ \p _{z''} \delta (z''-z') 
\underbrace{\delta ^{\nu \mu} [ -i k^\nu A^\mu + i A^\nu k^\mu ]}_{=0} \e \]
\be+ \p _{z''} X^\nu [ -ik^\nu A^\mu + ik^\mu A^\nu] ik^\mu \delta (z''-z')
\ee

The second term $\frac{\p L}{\p X^\mu(z)}\frac{\p L}{\p X^\mu (z')} $ becomes
\[\int dz'' \int dk [-A^\mu (k)  ik^\nu  + ik^\mu A^\nu ] \p _{z''} X^\nu (z'')\delta (z-z'') e^{ikX(z'')}\]
\[
\int dz''' \int dk [-A^\mu (k)  ik^\nu  + ik^\mu A^\nu ] \p _{z'''} X^\nu (z''')\delta (z'-z''') e^{ikX(z''')}
\]
Thus putting everything together we get
\[
\int dz~\int dz'~\dot G(z-z') \{\int dk [-A^\mu (k)  ik^\nu  + ik^\mu A^\nu ] ik^\mu \p _{z'} X^\nu (z')\delta (z-z') \e
\]
\be \label{Max}
+ \int dk \int dk' [-i k^{[\rho}A^{\mu ]}  ] [ -i{k'}^{[\sigma  } A^{\mu ]} ]\p _z X^\rho (z) \e \p _{z'}X^\sigma (z') e^{ik' X(z')}\}
\ee
The first
term is the usual Maxwell equation of motion that one should obtain at the linearized level in the beta function.
On performing the OPE in the second term (this assumes analyticity of $\dot G(z-z')$) we can rexpress
as a sum of vertex operators for the various modes, exactly as in the case of the tachyon, above.  

Notice that both terms are  gauge invariant.
From our experience with loop variables we see immediately
that this is because of the integral over $z,z'$ which allows integration by parts. We also see that  for the higher modes one need not expect full gauge invariance. In this fomalism the gauge invariance due to $L_{-1}$ is usually present - this is simply the freedom to add total divergences in $z$, which we have - as in the case of the photon. For the higher
gauge invariances due to $L_{-2}, L_{-3}...$ we need some additional variables. So in this formalism as it stands
, the equations are not invariant under the higher gauge transformations. In the next section we will
use the loop variable approach to address this problem of making the ERG  invariant under the full set of gauge transformations.

\section{Gauge Invariant ERG}
The gauge invariant construction involves writing the action in terms of the covariantized loop variable
\cite{BSLV,BSREV}
\be  \label{LV}
\int [{\cal D} k_n(t) dx_n]\gvks \Psi [k_n(t) ]
\ee

It is therefore useful to first redo the analysis of the previous section in the loop variable 
 formalism where we will use
the loop variable without covariantizing. This should reproduce the results of the previous section.
\be
\int {\cal D} k_n(t)\GVKS \Psi [k_n(t)]
\ee
Here $\tY _n = \frac{1}{(n-1)!}{\p ^n X\over \p t ^n}$.

\subsection{ERG in the Loop Variable Formalism - Gauge Fixed Case}

Let us consider for concreteness the vector field (level 1). We assume that
\[
\int [\prod _{n=1,2...}dk_n] \kim (t_1)\Psi [k_n] = A^\mu (k_0 (t_1))
\]
We also assume that
\be   \label{AA}
\int [\prod _{n=1,2...}dk_n] \kim (t_1)\kin (t_2)  \Psi [k_n] = A^\mu (k_0(t_1))A^\mu (\ko (t_2))  
\ee
and similarly for all products of $\kim (t_i)$. Furthermore
\[
\int [\prod _{n=1,2...}dk_n] k_i (t_1) k_j (t_2)  \Psi [k_n] = 0 ~~~~\forall i,j >1
\]  

which is equivalent to saying that we are setting the higher spin massive fields to zero.

In this notation we can write
\[
e^{i\int dt L[X(t)]} \equiv e^{i\int dt~\int dk~A_\mu (k)\e \p _t X^\mu (t)} = \int {\cal D} k_n(t)\GVKS \Psi [k_n(t)]
\]

To verify the correctness of this equation expand the LHS in powers of $A_\mu$. The linear term
is just $\int dt L[X(t)]$.  On the RHS the first term gives 
\[
\int {\cal D} k_n(t) i \int dt_1 \kim (t_1) \tY _1 (t_1) e^{i\int dt' k_0(t') X(t')} \Psi [k_n(t)]
\]
Using
\be    \label{Ak}
\langle \kim (t_1) \kon (t_2) \rangle = \delta (t_1-t_2) A^\mu (k_0(t_1)) k_0 ^\nu (t_1) 
\ee 
we see that the RHS becomes
\[
i \int dt_1 \int dk_0(t_1) A^\mu (k_0 (t_1)) \tY _1 (t_1) e^{i k_0(t_1) X(t_1)}
\]
which is the same as the LHS.

Let us go to the next order.  LHS gives:
\[ \hf
\int dt_1~ \int dt_2 \int dk_1 A^\mu (k_1) \p _{t_1}X^\mu (t_1) e^{ik_1X(t_1)}
 \int dk_2 ~A^\nu (k_2) \p _{t_2} X^\nu (t_2) e^{ik_2X(t_2)}
\]
In the loop variable expression we consider:
\[ \hf
\int {\cal D} k_n(t) i \int dt_1 ~\kim (t_1) \tY _1 ^\mu (t_1) \int dt_2~\kin (t_2) \tY _1 ^\nu (t_2)e^{i\int dt' k_0(t') X(t')} \Psi [k_n(t)]
\]
Using (\ref{AA}) and (\ref{Ak}) we see that
\[
=\hf \int dk_0(t_1) \int dk_0(t_2)~\int dt_1 A^\mu (k_0 (t_1)) \tY _1 ^\mu (t_1) e^{ik_0(t_1)}
\int dt_2 A^\nu (k_0 (t_2)) \tY _1 ^\nu (t_2)e^{ik_0(t_2)}
\]
which agrees with the LHS.

Let us now work out the ERG in this formalism:
\[
\frac{\delta ^2}{\delta X ^\mu (t') \delta X_\mu (t)} e^{i\int dt_1 [k_1(t_1) \p _{t_1} X(t_1) + k_0 (t_1) X(t_1)]}
\]
\[=
i\int dt_1~ \delta ^{\mu \nu}[\kim (t_1) \p _{t_1} \delta (t'-t_1) + \kom (t_1) \delta (t'-t_1)]
i\int dt_2~ [\kin (t_2) \p _{t_2} \delta (t-t_2) + \kon (t_2) \delta (t-t_2)]
\]
\[
 e^{i\int dt_3 [k_1(t_3) \p _{t_3} X(t_3) + k_0 (t_3) X(t_3)]}
\]
Let us concentrate on the part that is linear in $\kim$ (and therefore $A^\mu$). 

\[
\delta _{\mu \nu} 2i \int dt_1 \kim (t_1) \p _{t_1}\delta (t'-t_1) \int dt_2~ \kon (t_2) \delta (t- t_2) 
 e^{i\int dt_3  k_0 (t_3) X(t_3)}
\]
\[+ \delta _{\mu \nu} i\int dt_1 ~i\int dt_2~\kom (t_1) \kon (t_2) \delta (t'-t_1)\delta (t-t_2)
i\int dt_3~k_1^\rho (t_3) \p _{t_3} X^\rho (t_3)  e^{i\int dt_3  k_0 (t_3) X(t_3)}
\]
\[ = \delta _{\mu \nu}[2i\int dt_1 \kim (t_1) \kon (t_1) \p _{t_1} \delta (t-t_1) \delta (t'-t_1) e^{ik_0(t_1)X(t_1)}
\]
\[+ i \int dt_3 \kom (t_3) \kon (t_3) \delta (t-t_3) \delta (t'-t_3) k_1^\rho (t_3) \p _{t_3}X^\rho (t_3) e^{ik_0(t_3)X(t_3)}]
\]
Noting that $\p _{t_1} \delta (t-t_1) \delta (t'-t_1) = \delta (t-t_1) \p _{t_1}\delta (t'-t_1) =\hf \p _{t_1}
(\delta (t-t_1) \delta (t'-t_1))$ and integrating by parts,
\[ = \delta _{\mu \nu} i\int dt_1~[- \kim \kon ik_0 ^\rho (t_1) + \kom \kon k_1^\rho (t_1)]\p_{t_1}X^\rho (t_1)
e^{ik_0X(t_1)}
\]
This is just Maxwell's equation.

The part that has two $k_1$'s has three terms:
\[ (a)~~~\delta _{\mu \nu} i\int dt_1 \kim (t_1) \p _{t_1}\delta (t-t_1) i\int dt_2~ \kin (t_2) \p _{t_2} \delta (t'-t_2)
e^{i\int dt_3 k_0(t_3)X(t_3)}
\] 
\[(b)~~~2\delta _{\mu \nu} i\int dt_1 \kim (t_1) \p _{t_1}\delta (t-t_1)i\int dt_2 \kon (t_2) \delta (t'-t_2)
i\int dt_3 \ki (t_3) \p _{t_3}X(t_3)e^{i\int dt_4 k_0(t_4)X(t_4)}
\]
\[(c)~~~\delta _{\mu \nu}   i\int dt_1 \kom (t_1) \delta (t-t_1)i\int dt_2 \kon (t_2) \delta (t-t_2)   
  \ki (t_1) \p _{t_1}X (t_1)\ki (t_2) \p _{t_2}   X (t_2)           e^{i\int dt_3 k_0(t_3)X(t_3)}
\]

Integrating by parts on $t_1$ and $t_2$ in (a)
\[ \delta _{\mu \nu} i\kim (t) k_0^\rho \p _t X^\rho (t) i\kin (t') k_0^\sigma \p _{t'} X^\sigma (t') 
e^{i \int dt _4 k_0(t_4) X(t_4)}
\]

Similarly one can simplify (b) and (c) to get finally:
\[
\delta _{\mu \nu} k_1^{[\mu}(t) k_0^{\rho ]}(t) \p _t X^\rho 
 k_1^{[\nu}(t') k_0^{\sigma ]}(t') \p _{t'} X^\sigma e^{i[k_0 (t) X(t) )+ k_0 (t') X(t')]}
\]

One also sees that more than two $k_1$'s cannot contribute to the equation: there is
 one $k_1 (t)$ and one $k_1(t')$. Additional $k_1(t)$ would introduce massive fields
which we have set to zero. The rest of the terms in the exponental $e^{i\int dt \ki (t) \p X(t)}$
constitute an overall multiplicative factor (involving $A^\mu (X)$) in the ERG equation and we need
not worry about it. This is exactly as happens in the conventional field theory formalism.

These results can now be compared with (\ref{Max}) and are seen (on using (\ref{AA})) to agree.

\subsection{ERG - Gauge Invariant Loop Variable Formalism}

We can proceed to do repeat the above calculation in the gauge invariant calculation. The main difference
is that instead of integrating by parts on $t$ we integrate by parts on  an infinite number of variables, 
$x_n$ - but these are global
i.e. not $x_n(t)$. This implies that vertex operators at all locations participate in the ERG equations, which
therefore, are no longer quadratic in fields. Another important difference is that right from the beginning, all  vertex operators are  Taylor expanded about one point in the world sheet. This is equivalent to Taylor expanding
$\dot G(u)$. The Gaussian fall off at distances of order the cutoff, $a$, is thus not seen in this power series
expansion. The upshot is that one needs an IR cutoff $R$ and the dimensionless ratio $R/a$ enters in
all the equations. 

Our starting point in (\ref{LV}).  We act with 
$\int dt \int dt' \dot G(t-t')\frac{\delta ^2}{\delta X^\mu (t)\delta X_\mu (t')}$ on 
$\int [{\cal D} k_n(t) dx_n]\gvks \Psi [k_n(t) ]$.
We use \cite{BSLV,BSREV} \[ Y (t) = X(t) + \al _1 \p  X(t) + \al _2 \pp X(t) + \hf \al _3 \ppp X(t) +...+{\p ^n X(t)\over (n-1)!}+...
\] 
where \[ \sum _n \aln t^{-n} = e^{\sum _n \xn t^{-n}} \]
and \[ \frac{\delta Y(t)}{\delta X(t')}= \sum _{n=0}^{\infty} \frac{\al _n \p _t^n \delta (t-t')}{(n-1)!} \]

The loop variable is however rewritten after {\em first} Taylor expanding about the point $t=0$ and {\em then} covariantising, as
$e^{i\sum _n \int dt \bar k_n (t) \yn (0) }$.
Thus we get
\[
\int dt \int dt' ~ \dot G(t-t') \int \prod _{r \ge 1}[{\cal D} k_r(t) dx_r]  
 i\sum _n  \int dt_1 
\bar k_n (t_1) \dsn [ \sum _p \frac{\al _p \p ^p _{t_2}\delta (t-t_2)}{(p-1)!}|_{t_2=0}]
\]
\[
 i\sum _m  \int dt_3 
\bar k_m (t_3) \dsm [ \sum _q \frac{\al _q \p ^q _{t_4}\delta (t'-t_4)}{(q-1)!}|_{t_4=0}]
 e^{i\sum _r \int dt_5 \bar k_r (t_5) Y_r (0) }     \Psi [k_n(t) ]
\]
Rewrite $\frac{\p}{\p t_2}$ as $-\frac{\p}{\p t }$ and integrate by parts on $t$ to act on $\dot G(t-t')$, and do the
same for $t_4$ (rewrite in terms of $t'$). We get:
\[ \int\prod _{s\ge 1} [{\cal D} k_s(t) dx_s]\int dt \int dt'~\sum _{n,m \ge 0}\{\dsn [ \sum _p \frac{\al _p \p ^p _{t} }{(p-1)!}] \dsm [ \sum _q \frac{\al _q \p ^q _{t'}}{(q-1)!} ]\dot G(t-t')\}\]\[
\int dt_1 \int dt_3 ~\bar k_n(t_1) . \bar k_m (t_3) \delta (t) \delta (t') e^{i\sum _r \int dt_5 \bar k_r (t_5) Y_r (0) }     \Psi [k_n(t) ]
\] 
\[=\int\prod _{s\ge 1} [{\cal D} k_s(t) dx_s]~\sum _{n,m \ge 0}\underbrace{\{\dsn [ \sum _p \frac{\al _p \p ^p _{t} }{(p-1)!}] \dsm [ \sum _q \frac{\al _q \p ^q _{t'}}{(q-1)!} ]\dot G(t-t')|_{t=t'=0}\}}_{\hf (\dsnm -\dsq )\Sigma (0)}\]\[
\int dt_1\int dt_3~ \bar k_n(t_1) . \bar k_m (t_3)  e^{i\sum _r \int dt_5 \bar k_r (t_5) Y_r (0) }     \Psi [k_n(t) ]
\]

We assume that when $n=0$ $\dsn =1$.
Implementing the delta functions in $t,t'$ we see the Taylor expansion of $\dot G(u)$ mentioned above.
The object $\Sigma$ can be identified with the generalized Liouville mode introduced in earlier papers
on loop variables. There the EOM were obtained by varying w.r.t $\Sigma$. Here we have two options.
$\Sigma$ is something that can be evaluated (in principle) as a function of $\xn$. We can then take this equation
as it stands, and evaluate it at say, $\xn =0$. This will then reproduce the gauge fixed equation derived
in the last section. In this form the equation will be quadratic in fields, because the higher order 
terms just factorize. We saw this in the example where we had only $A^\mu$.  In the more general case,
massive modes will be involved, but the factorization will still be true. However this equation
is not gauge invariant. The second option is to integrate by parts on $\xn$ so that there are no derivatives
on $\Sigma$. The coefficient of $\Sigma$ is gauge invariant. We can use this as the EOM. This is what
is done in the gauge invariant loop variable formalism. However the equations no longer factorize. They reduce
to the sum of two terms, each of which factorizes. This can be seen as follows:

Schematically the exponential $\gvks$ can be written as $E(T)E(T')$ where $T$ corresponds to either
$t_1$ or $t_3$ in the above expression and $T'$ represents any other value of $t$. Thus in terms of fields
there will be a factorization: $\langle E(T') \rangle$ is an overall multiplication factor. E(T) introduces
higher modes into the equation, but as there are only two points $t_1$ and $t_3$, it is always
a product of two fields and so the  equations are quadratic.  Now consider the effect of integration
by parts: We get terms of the form $\dsn (E(t) E(T')) = (\dsn E(T))E(T') + E(T) \dsn E(T')$. Each term can be
seen to factorize, but the sum clearly will not be factorizable.  Since it is only the sum that is gauge invariant, we no
longer get a quadratic equation.

The RHS of the gauge invariant equation obtained is (LHS is just the beta function of each field):  

\[\int\prod _{s\ge 1} [{\cal D} k_s(t) dx_s]~\sum _{n,m \ge 0} \int dt_1 \int dt_3 ~\bar k_n(t_1) . \bar k_m (t_3) 
\]
\[ \hf ( \dsnm + \dsq ) [e^{i\sum _r \int dt_5 \bar k_r (t_5) Y_r (0) }]     \Psi [k_n(t) ]
\]
After differentiation, one can set $\xn =0$ to evaluate the expressions. The LHS is an expansion in 
$\yn$ and so is the RHS. Thus matching coefficients we get an infinite number of equations - each equation
defines a beta function. If we set LHS=RHS=0 we get the condition for the fixed point and this is the
EOM for the fields of the string. 

Let us work out the electromagnetic case worked out earlier:
\[ [\Sigma (0) k_0.k_0  + \dsi \Sigma (0) k_1.k_0] \gvks\]
\[=\Sigma (0)(k_0.k_0  -   k_1.k_0\dsi ) \gvks = \Sigma (0) (k_0.k_0 i\kim -   k_1.k_0 i\kom )Y_1 ^\mu e^{ik_o.Y} +... \]
where the three dots indicate  terms involving other vertex operators. On the LHS
is $\langle i\dot \kim Y_1^\mu e^{ik_0Y}+...\rangle$.   

Gauge invariance of the equations follow exactly as in the usual loop variable formalism. The generalized tracelessness  constraint \[
\langle \int dt \int dt_1 \int dt_2 \la _p (t) \bar k_n (t_1) . \bar k_m (t_2) .....\rangle =0~~~\forall n,m>0
\]
 is also required (as before).   
 
The dimensional reduction with mass has to be done exactly as in the usual case. Thus $q_n (t)$ is the generalized loop variable  momentum in the 27th dimension. $q_0^2$ is to be set equal to the engineering dimension of the operator. We do not need $\bar q _n (t)$ because it is assumed that the long distance part 
of the Green's function is zero for the 27th coordinate - so $X^{26}$ has no $t$-dependence, so  $\bar q_n(t) = q_n(t)$. 

\section{Conclusion}

We have written down an exact renormalization group equation for the world sheet theory describing
a general open string background. These equations are valid for finite cutoff. Indeed in the limit
$R\rightarrow \infty$, the cutoff parameter $a$ does not enter the ERG. This means the finite cutoff
RG equations  - or the theory on a lattice with finite spacing - is the same as with $a=0$. An action
with this property  has been described as a "perfect" action \cite{PH} - and also earlier similar ideas were introduced under the name of  "improved" action \cite{KS}.  Furthermore  if we tune the parameters so that the beta functions are set to zero, then the resulting action is conformally invariant even if $a \neq 0$! 

Furthermore using the loop variable formalism it is possible to make these equations gauge invariant.
Gauge invariance usually ensures that the space-time theory is consistent. It is therefore
a good check on the procedure. This takes the place of the usual checks such as BRST. Since there is a lot
of freedom in the world sheet action, there is the possibility that this procedure can have a certain
background independence, in that all backgrounds are on an equivalent footing. This was exploited
in \cite{BSA} to get a gauge invariant space time action for massive higher spin modes in AdS space time.
One should be able to do this in a more general way using the ERG. This remains to be investigated.

 However there is one new feature -  at least in the way gauge invariance is achieved here, the parameter $R/a$ enters the equation.  
This can be traced to the Taylor expansion of all vertex operators about one common point. 
Of course the value of the parameter is arbitrary. If one were to fix gauge, one could resum the series and take the
limit $R\rightarrow \infty$. In the expanded form it is not possible to take $R\rightarrow \infty$. The limit $R/a \rightarrow 0$
seems well defined - although it may not be physically reasonable. It may be that critical information is lost in this limit, thus  invalidating it. This remains to be explored. It is worth pointing out that string field
theory (BRST, light cone ) also  has some similar parameter, though for a given formulation it is a fixed number. 

Finally on a speculative level, the role of a finite cutoff in {\em space-time} (as against world sheet) 
was discussed in \cite{BSLV}.  
The speculation was that string theory effectively imposes a finite space-time cutoff, and the large gauge symmetry - a generalized RG - of string theory makes the details of the cutoff unimportant, i.e. physically unobservable.  Thus string theory should then be an example of a "perfect" (in the RG sense)
space-time action. 

\vspace{1cm}
{\bf Acknowledgements:} I would like to thank P. Ray for bringing \cite{JC} to my attention and for some useful discussions. I also thank G. Date for a useful discussion.

\end{document}